\documentclass[pss,fleqn,embeddedheads]{w-art}
\usepackage{times}
\usepackage{w-thm}
\usepackage[]{graphicx}
\usepackage{amssymb}

\newcommand{\bra}[1]{\langle #1|}
\newcommand{\ket}[1]{|#1\rangle}

\newcommand{\boldgreek}[1]{\ensuremath{\mbox{\boldmath$#1$}}}

\begin{document}
\DOIsuffix{theDOIsuffix}
\Volume{XX}
\Issue{1}
\Month{01}
\Year{2003}
\pagespan{3}{}
\Receiveddate{21 August 2005}
\Reviseddate{27 August 2005}
\Accepteddate{}
\Dateposted{$Revision: 1.22 $, compiled \today}



\title[Scaling, the Energy-driven MIT and the Thermoelectric Power]{Scaling at the Energy-driven Metal-Insulator Transition and the Thermoelectric Power}


\author[Alexander Croy]{Alexander Croy\footnote{Corresponding
     author: e-mail: {\sf a.croy@warwick.ac.uk}, Phone: +44\,2476\,574\,309,
     Fax: +44\,7876\,858\,246}\inst{1}\inst{2}}
\address[\inst{1}]{Institut f\"ur Physik, Technische Universit\"at, 09107 Chemnitz, Germany}
\author[Rudolf A.\ R\"{o}mer]{Rudolf A.\ R\"{o}mer\inst{2}}
\address[\inst{2}]{Physics Department \& Centre for Scientific Computing,
                   University of Warwick, Coventry CV4 7AL, U.K.}

\begin{abstract}
  The electronic properties of disordered systems at the Anderson
  metal-insulator transition (MIT) have been the subject of intense
  study for several decades. Thermoelectric properties at the MIT, such
  as thermopower and thermal conductivity, however, have been relatively
  neglected. Using the recursive Green's function method
  and the Chester-Thellung-Kubo-Greenwood formalism, we 
  calculate numerically the low temperature behaviour of all kinetic
  coefficients $L_{ij}$.  From these we can deduce for example the electrical
  conductivity $\sigma$ and the thermopower $S$ at finite temperatures. 
  Here we present results for the case of completely coherent transport in cubic 3D systems.
\end{abstract}
\maketitle                   







%

\section{Introduction}
\label{SecIntroduction}
The Anderson model \cite{And58} is widely used to investigate the
phenomenon of localisation in disordered materials. Especially the
possibility of a quantum phase transition driven by disorder from an
insulating phase, where all states are localised, to a metallic phase
with extended states has lead to extensive analytical and numerical
investigations of the critical properties of this metal-insulator
transition (MIT) \cite{KraM93}.

The one-parameter scaling theory plays a crucial role in understanding
the MIT \cite{AbrALR79}. It is based on an ansatz interpolating between metallic
and insulating regimes \cite{LeeR85}. So far, scaling has been demonstrated to an 
astonishing degree by numerical studies of the Anderson model \cite{SleO99a}.
However, most studies focused on scaling of the localisation length and
the conductance at the disorder-driven MIT in the vicinity of the band centre.

In the present paper, we will show numerical results for cubes of volume
$L^3$ which are consistent with scaling also at the energy-driven
transition across the mobility edge $E_{\rm c}$ near the band tails. In
particular, we find that at $T=0$ the d.c.\ conductivity $\sigma(E)$ close to 
$E_{\rm c}$ is well described by the power-law \cite{LeeR85,Weg76}
\begin{equation}
  \sigma(E) = \left\{
    \begin{array}{cl}
      \sigma_0 \left|1-\frac{E}{E_{\rm c}} \right|^\nu & |E| < E_{\rm c}, \\
      0,                                             & |E| > E_{\rm c}.
    \end{array}\right.\;
  \label{powerlawsigma}
\end{equation}

Moreover, knowing the explicit energy and system size dependence of the
conductance $g= \sigma L$ allows us to calculate the low-temperature dependence
of thermoelectric properties such as thermopower, Lorenz number and thermal
conductivity in the fully quantum coherent regime. For these, similarly detailed
theoretical investigations of scaling at the MIT are still lacking. Here we
present an approach based on the recursive Green's function technique \cite{Mac80}.

\section{Thermoelectric Transport Properties}
Thermoelectric transport properties describe the response of a material
to the presence of a small external electric field and a small
temperature gradient $\boldgreek{\nabla} T$. E.g., the induced ({\it
  thermo-}){\it electric field} ${\textbf{\em E}}$ in the Seebeck effect is directly
proportional to $\boldgreek{\nabla} T$ and the constant of
proportionality is called {\it thermopower} $S$.
In general, the dependency of the electric (${\textbf{\em j}}_{\rm e}$) and thermal
currents (${\textbf{\em j}}_{\rm q}$) on a electric field and a temperature
gradient up to linear order are \cite{Cal85}
\begin{equation}
    {\textbf{\em j}}_e = \frac{1}{|e|} \left( |e|\,L_{11}{\bf {\bf \textit {E}}} - L_{12}\frac{\boldgreek{\nabla} T}{T} \right) \qquad\text{and}\qquad
    {\textbf{\em j}}_q = \frac{1}{|e|} \left( |e|\,L_{21}{\bf {\bf \textit {E}}} - L_{22}\frac{\boldgreek{\nabla} T}{T} \right)\;. \label{Curr}
\end{equation}
Here $e$ is the electron charge. This defines the {\it kinetic
  coefficients} $L_{ij}\;(i,j=1,2)$. Additionally, for zero
magnetic field, the Onsager relation states that $L_{21}=L_{12}$
\cite{Cal85}.
Using the definitions of the thermoelectric properties \cite{Cal85} and
Eq.\ (\ref{Curr}) one can express $\sigma$ and $S$ in terms of the
kinetic coefficients as
\begin{equation}
    \sigma = L_{11 } \qquad\text{and}\qquad   S = \frac{L_{12}}{|e| T L_{11}}\;.
    \label{ThermoProp}
\end{equation}
%
Assuming elastic scattering of independent electrons by static impurities or by
lattice vibrations one can obtain the kinetic coefficients from the
Chester-Thellung-Kubo-Greenwood (CTKG) formulation \cite{CheT61,Gre58},
\begin{equation}
    L_{ij}(T) = (-1)^{i+j} \int\limits_{-\infty}^{\infty}\;
    \sigma(E) [E-\mu(T)]^{(i+j-2)} \left[- \frac{\partial f(E,\mu,T)}{\partial E} \right] dE\;,
    \quad (i,j=1,2)\,,
    \label{CTKG}
\end{equation}
where $\mu(T)$ is the chemical potential and $f(E,\mu,T)$ is the usual Fermi function.
%
\section{Model and Numerical Method}
The Anderson model of localisation is based upon a tight-binding
Hamiltonian in site representation
\begin{equation}
    \mathcal{H} = \sum_{i} \varepsilon_i \ket{i}\bra{i} - \sum_{i \neq j} t_{ij}\,\ket{i}\bra{j}\;,
    \label{AndersonHam}
\end{equation}
where $\ket{i}$ is a localised state at site $i$ and $t_{ij}=t$ are the
hopping parameters, restricted as usual to nearest neighbours. The
on-site potentials $\varepsilon_i \in [-W/2,W/2]$ are uniformly distributed random numbers.

For calculating the energy dependence of the two-point conductance $G_2=e^2/h\,g_2$ we use
the recursive Green's function method \cite{Mac80}. This method is based
on the Kubo-Greenwood formula for the linear response regime and non-interacting
electrons \cite{Gre58}.  In order to obtain correct results for purely elastic
scattering, metallic leads were attached at both ends of the system. In this case
the formalism given above is equivalent \cite{Nik01} to the Landauer-B\"uttiker
formulation \cite{Lan70}. To get the conductance $g$ of the disordered region only, 
we have to subtract the contact resistance due to the leads. This gives 
${1}/{g} = {1}/{g_2} - {1}/{N}$.  Here $N=N(E)$ is the number of propagating 
channels at the Fermi energy $E$ \cite{But88b}.
An additional complication arises because the number of transmitting
lead-modes drops to zero outside the lead-energy band, whereas the
disordered system still has states at these energies. This mismatch
gives rise to a lead-induced reduction in $g$. The problem
can be overcome by shifting the energy of the disordered region while
keeping the Fermi energy in the leads in the lead-band centre.  This is
equivalent to applying a gate voltage to the disordered region and
sweeping it --- a technique similar to MOSFET experiments.
%
\section{Conductance Scaling and D.C. Conductivity}\label{SecConductivity}
We set the disorder strength to $W/t = 12$ and impose fixed boundary
conditions in the transverse direction \cite{SleMO01}. For each
combination of Fermi energy and system size we generate an ensemble of 
$10000$ samples (except for $L=19$ and $L=21$, where $4000$ and $2000$ were generated) 
and examine the size-dependence of the average and the 
typical conductance, $\langle g \rangle$ and $\langle \ln g \rangle$, respectively. 
We find that for $E_{\rm F}/t \le -8.2$ the typical conductance is proportional to the 
system size $L$ and the constant of proportionality is negative. This corresponds to 
an exponential decay of the conductance with increasing $L$ and
is characteristic for {\it insulating} behaviour. For $E_{\rm F}/t$ being larger than
$-8.05$, $\langle g \rangle$ is proportional to $L$. This indicates
the {\it metallic} regime.
\begin{figure}[h]
    \center
    \begin{minipage}[b]{.45\textwidth}
        \includegraphics[width=\textwidth]{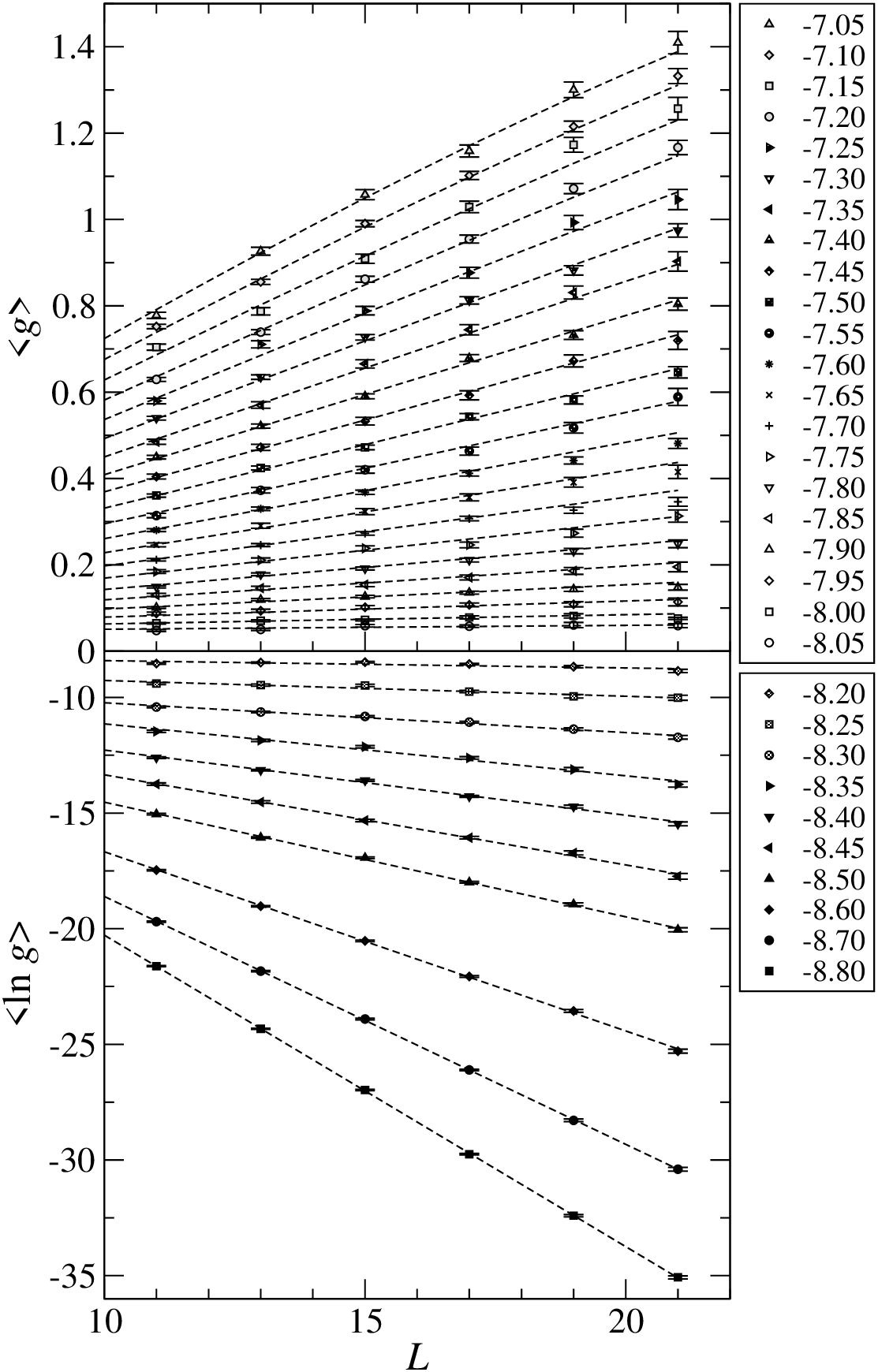}
    \end{minipage}\hfill
    \begin{minipage}[b]{.52\textwidth}
      \center
      \includegraphics[width=.95\textwidth]{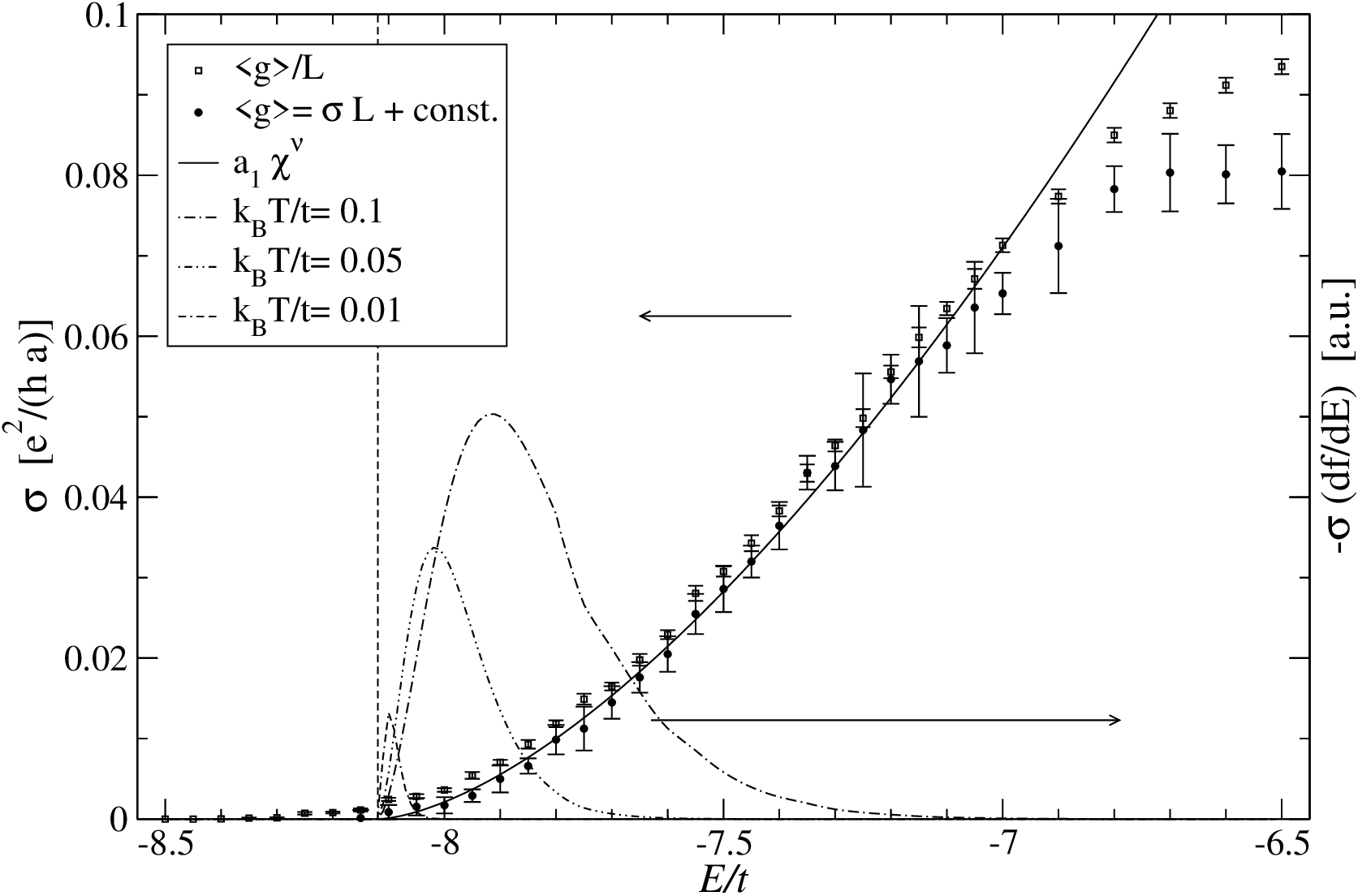}
        \caption{
          (top) Conductivity $\sigma$ vs energy computed from $\langle g\rangle/L$
          ($\square$), a linear fit with $\langle g\rangle = \sigma L +{\rm const.}$
          ($\bullet$) and a scaling according to Eq.\ (\ref{eq-scaling-AC}) 
	  (solid line). The dashed line indicates $E_{\rm c}/t= -8.12$. Error bars 
	  of $\langle g\rangle/L$ represent the error-of-mean obtained from an ensemble 
	  average. Also shown is the integrand of Eq.\ (\ref{CTKG}) for $i=j=1$
	  and three different temperatures (dash-dotted).}
        \label{fig:Conductivity}
        \caption{
          (left) System size dependence of the 4-point conductance
          averages $\langle g \rangle$ and $\langle \ln g \rangle$ for
          $W/t=12$ and Fermi energies are given in the legend. Error bars
	  obtained from ensemble average. The dashed lines indicate the fit 
	  results to Eq.\ (\ref{eq-scaling-AC}) and a linear function to 
	  $\langle g\rangle$ and $\langle \ln g \rangle$, respectively.}
        \label{fig:Scaling}
    \end{minipage}
\end{figure}

Following Ref.\ \cite{SleMO01} we fit the data in the respective regimes
to a standard scaling form, i.e.\ we assume 
\begin{equation}
    \langle g \rangle = F(\chi L^{1/\nu})
    \qquad\text{and}\qquad 
    \langle \ln g \rangle = F(\chi L^{1/\nu}),
    \label{SlevinScaling}
\end{equation}
where $\chi$ is a relevant scaling variable. The results for the critical
exponent and the mobility edge are given in Table\ \ref{tab:FSSResults}. The 
obtained values for both types of averages, $\langle g\rangle$ and
$\langle \ln g\rangle$, are consistent. The average value of $\nu= 1.59 \pm 0.18$
is in accordance with results for conductance scaling at $E_{\rm F}/t=0.5$ and
transfer-matrix calculations \cite{SleO99a,SleMO01}.

In order to obtain the anticipated power-law form for the conductivity
$\sigma(E)$ in the critical regime, we need to assume the following
scaling law for the conductance,
\begin{equation}
        \langle g\rangle = f( \chi^\nu L )\;.
\label{eq-scaling-AC}
\end{equation}
Then we expand $f$ as a Taylor series up to order $n_{\rm R}$ and $\chi$ in terms of the
parameter $\varepsilon = (E_{\rm c}-E)/E_{\rm c}$ up to order $m_{\rm R}$. This procedure
gives
\begin{equation}
        f( \chi^\nu L ) = \sum\limits^{n_{\rm R}}_{m=0} a_m \left( \chi^\nu L \right)^m \qquad\text{and}\qquad
        \chi(\varepsilon) = \sum\limits^{m_{\rm R}}_{n=1} b_n \varepsilon^n \;.
\end{equation}
The best fit is determined by minimising the $\chi^2$ statistic. Using
$n_{\rm R}=3$ and $m_{\rm R}=2$ we obtain for the critical values, $\nu = 1.58 \pm
0.18$ and $E_{\rm c}/t = -8.12 \pm 0.03$. These values are consistent with 
our previous fits.  The $m=1$ term $a_1 \chi^\nu$ in $f$ corresponds to the
conductivity close to the MIT. To estimate the quality of this
procedure we calculate the conductivity from the slope of a linear fit
to $\langle g\rangle$ throughout the metallic regime, and also from the
ratio $\langle g\rangle/L$. The resulting conductivities are shown
in Fig.\ \ref{fig:Conductivity}.
\begin{table}
  \center
  \begin{tabular}{lcccccc}
    average                & $E_{\rm min}/t$ &  $E_{\rm max}/t$ & $n_{\rm R}$ & $m_{\rm R}$ & $\nu$            & $E_{\rm c}/t$ \\
    \hline
    \hline
    $\langle g\rangle$     & $-8.2$          &  $-7.4$          & $3$         & $2$         & $1.60 \pm 0.18$  & $-8.14  \pm 0.02$ \\
    $\langle \ln g\rangle$ & $-8.8$          &  $-7.85$         & $3$         & $2$         & $1.58 \pm 0.06$  & $-8.185 \pm 0.012$
  \end{tabular}
  \caption{Best fit estimates of the critical exponent and the mobility edge for both averages of $g$ using Eq.\ (\ref{SlevinScaling}). 
    The system sizes used were in both cases $L=11,13,15,17,19,21$.}
  \label{tab:FSSResults}
\end{table}
%
\section{Thermoelectric Transport}\label{SecTransport}
\begin{figure}[h]
    \center

    \begin{minipage}[b]{.49\textwidth}
        \includegraphics[width=\textwidth]{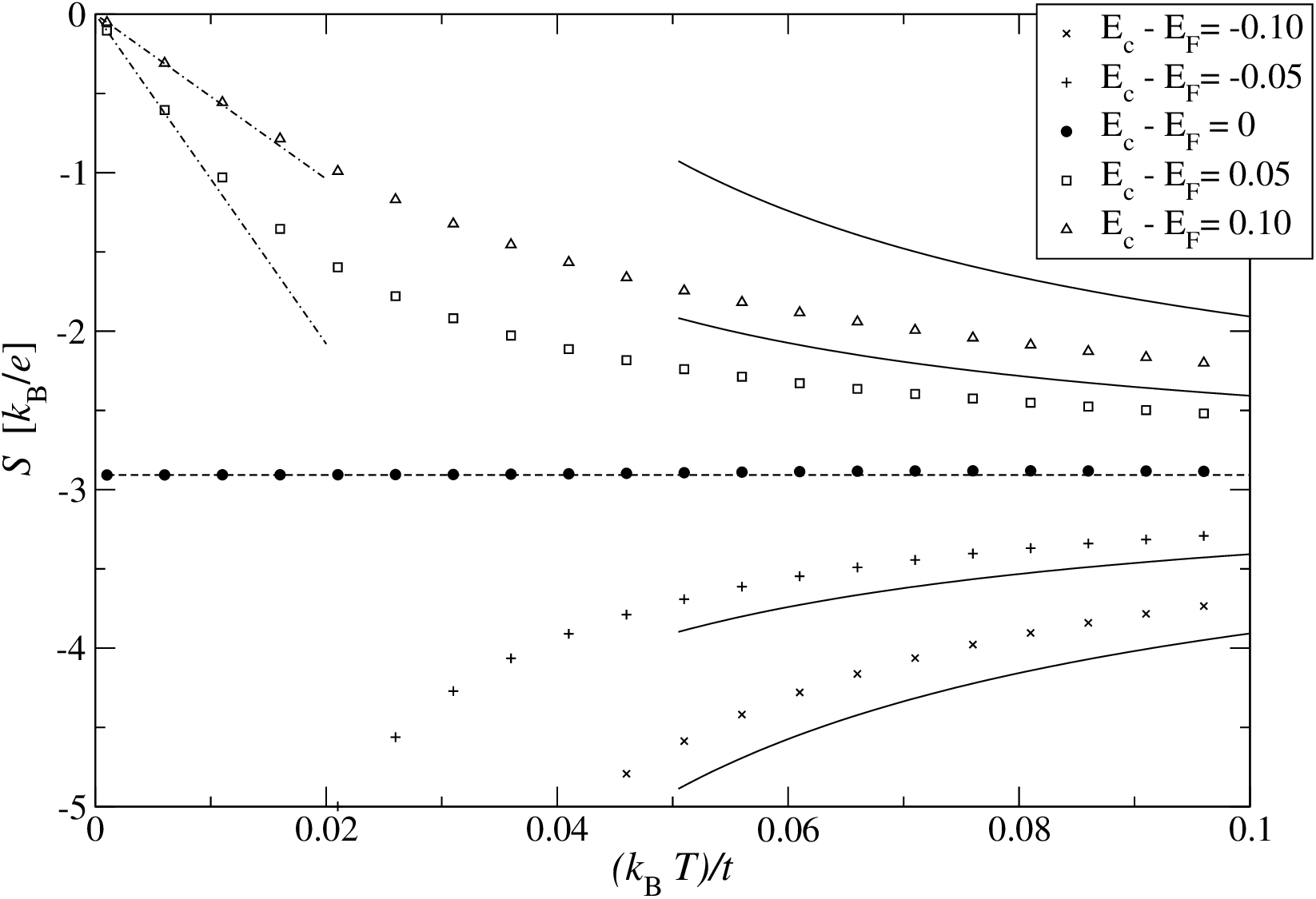}
    \end{minipage}\hfill
    \begin{minipage}[b]{.49\textwidth}
        \caption{Thermopower $S$ vs a dimensionless temperature parameter $k_{\rm B} T/t$. Empty symbols indicate
      {\it metallic}, filled symbols {\it critical} and stars {\it insulating} behaviour. The distance from
      $E_{\rm c}$ is given in the legend. Full lines are obtained from a high-temperature expansion 
      \cite{VilRS99a}. The dashed lines indicate $S$ using the Sommerfeld expansion \cite{VilRS99a}. In both 
      cases $\nu$ and $E_{\rm c}$ obtained from the fit to Eq.(\ref{eq-scaling-AC}) are used.}
        \label{fig:Thermopower}
    \end{minipage}
\end{figure}

Knowing the energy dependence of $\sigma$ it is straightforward to evaluate
Eq.\ (\ref{CTKG}) and to compute the temperature dependence of $\sigma$ and
$S$ using Eq.\ (\ref{ThermoProp}). To investigate the consequences of deviations
from the power-law behaviour of $\sigma(E)$ on $S(T,E_{\rm F})$ we
only use $\sigma=a_1 \chi^\nu$ for energies close to the MIT ($E_{\rm F}/t \le -7.8$). 
Otherwise an interpolation to $\sigma$ obtained from the numerical data is used.
The results for the thermopower together with the low- and high-temperature
expansions using only Eq.\ (\ref{powerlawsigma}) are shown in
Fig.\ \ref{fig:Thermopower}. Deviations from the pure power-law result
\cite{VilRS99a} accumulate above $k_{\rm B} T /t \approx 0.5$.

In Fig.\ \ref{fig:Conductivity} one can see that at this temperature the integrand in
Eq.\ (\ref{CTKG}) has already a significant contribution from the pure numerical
data. Nevertheless, the deviations of $S$ from the expansions are small even for
higher temperatures. This indicates a certain robustness of the results in Ref.\ 
\cite{VilRS99a}.
%
\begin{acknowledgement}
  We thank C.\ Sohrmann and M.\ Schreiber for useful discussions.
\end{acknowledgement}


\begin{thebibliography}{10}

\bibitem{And58}
P.~W. Anderson, Phys. Rev. {\bf 109},  1492  (1958).

\bibitem{KraM93}
B. Kramer and A. MacKinnon, Rep. Prog. Phys. {\bf 56},  1469  (1993).

\bibitem{AbrALR79}
E. Abrahams, P.~W. Anderson, D.~C. Licciardello, and T.~V. Ramakrishnan, Phys.
  Rev. Lett. {\bf 42},  673  (1979).

\bibitem{LeeR85}
P.~A. Lee and T.~V. Ramakrishnan, Rev. Mod. Phys. {\bf 57},  287  (1985).

\bibitem{SleO99a}
K. Slevin and T. Ohtsuki, Phys. Rev. Lett. {\bf 82},  382  (1999);
%
T. Ohtsuki, K. Slevin, and T. Kawarabayashi, {Ann. Phys. (Leipzig)} {\bf 8},
  655  (1999);
%
R.~A. {R\"{o}mer} and M. Schreiber,  in {\em The Anderson Transition and its
  Ramifications --- Localisation, Quantum Interference, and Interactions},
  edited by T. Brandes and S. Kettemann (Springer, Berlin, 2003),
  pp.\ 3--19.

\bibitem{Weg76}
F. Wegner, Z. Phys. B {\bf 25},  327  (1976).

\bibitem{Mac80}
A. MacKinnon, J. Phys.: Condens. Matter {\bf 13},  L1031  (1980).
%
---, Z. Phys. B {\bf 59},  385  (1985).

\bibitem{Cal85}
H.~B. Callen, {\em Thermodynamics and an Introduction to Thermostatistics}
  (John Wiley \& Sons, New York, 1985).


\bibitem{CheT61}
G.~V. Chester and A. Thellung, Proc. Phys. Soc. {\bf 77},  1005  (1961).

\bibitem{Gre58}
D.~A. Greenwood, Proc. Phys. Soc. {\bf 71},  585  (1958);
%
R. Kubo, J. Phys. Soc. Japan {\bf 12},  570  (1957).

\bibitem{Nik01}
B.~K. Nikoli\'{c}, Phys. Rev. B {\bf 64},  165303  (2001).

\bibitem{Lan70}
R. Landauer, Phil. Mag. {\bf 21},  863  (1970);
%
M. {B\"{u}ttiker}, Phys. Rev. Lett. {\bf 57},  1761  (1986).

\bibitem{But88b}
M. B\"{u}ttiker, Phys. Rev. B {\bf 38},  9375  (1988).

\bibitem{SleMO01}
K. Slevin, P. Marko\u{s}, and T. Ohtsuki, Phys. Rev. Lett. {\bf 86},  3594
  (2001).

\bibitem{VilRS99a}
C. Villagonzalo, R.~A. {R\"{o}mer}, and M. Schreiber, Eur. Phys. J. B {\bf 12},
   179  (1999).

\end{thebibliography}

\end{document}